\documentclass[aps,prd,twocolumn,showpacs,superscriptaddress]{revtex4}

\usepackage{graphicx}
\usepackage{color}
\usepackage{natbib}
\usepackage{epsfig}

\usepackage{amsmath,amssymb,amsfonts}

\begin{document}

\title{A unified picture of the post-merger dynamics and gravitational wave emission in neutron-star mergers}

\author{A.~Bauswein}
\affiliation{Department of Physics, Aristotle University of
  Thessaloniki, GR-54124 Thessaloniki, Greece}

\author{N.~Stergioulas}
\affiliation{Department of Physics, Aristotle University of
  Thessaloniki, GR-54124 Thessaloniki, Greece}

\date{\today} 

\begin{abstract}
We introduce a classification scheme of the post-merger dynamics and gravitational-wave emission in binary neutron star mergers, after identifying a new mechanism by which a secondary peak in the gravitational-wave spectrum is produced. It is caused by a spiral deformation, the pattern of which  rotates slower with respect to the double-core structure in the center of the remnant. This secondary peak is typically well separated in frequency from the secondary peak produced by a nonlinear interaction between a quadrupole and a quasi-radial oscillation. The new mechanism allows for an explanation of low-frequency modulations seen in a number of physical characteristics of the remnant, such as the central lapse function, the maximum  density and the separation between the two cores, but also in the gravitational-wave amplitude. We find empirical relations for both types of secondary peaks between their gravitational-wave frequency and the compactness of nonrotating individual neutron stars, that exist for fixed 
total binary masses. These findings are derived for equal-mass binaries without intrinsic neutron-star spin analyzing hydrodynamical simulations without magnetic field effects. Our classification scheme may form the basis for the construction of detailed gravitational-wave templates of the post-merger phase. We find that the quasi-radial oscillation frequency of the remnant decreases with the total binary mass. For a given merger event our classification scheme may allow to determine the proximity of the measured total binary mass to the threshold mass for prompt black-hole formation, which can, in turn, yield an estimate of the maximum neutron-star mass.

\end{abstract}


   \maketitle

\section{Introduction}
Neutron star (NS) mergers are strong emitters of gravitational waves (GWs) and thus among the prime targets for the upcoming GW detectors Advanced LIGO~\cite{2010CQGra..27h4006H} and Advanced Virgo~\cite{2006CQGra..23S.635A}. Future GW observations of such events~\cite{2010CQGra..27q3001A} could reveal the properties of high-density matter and NSs (see e.g.~\cite{2010CQGra..27k4002D,2011GReGr..43..409A,BaumgarteShapiro,2012LRR....15....8F,Rezzolla} for reviews). The merger will likely result in the formation of a differentially rotating, strongly oscillating remnant~\cite{1994PhRvD..50.6247Z,1996A&A...311..532R,2000A&A...360..171R,2002PhRvD..65j3005O,2002PThPh.107..265S,2004PhRvD..69l4036F,2005PhRvL..94t1101S,2005PhRvD..71h4021S,2006PhRvD..73f4027S,2007A&A...467..395O,2007PhRvL..99l1102O,2008PhRvD..77b4006A,2008PhRvD..78b4012L,2008PhRvD..78h4033B,2009PhRvD..80f4037K,2011PhRvD..83d4014G,2011MNRAS.418..427S,2011PhRvD..83l4008H,2011PhRvL.107e1102S,2012PhRvL.108a1101B,2012PhRvD..86f3001B,2012PhRvD..86f4032P,2013arXiv1306.4065R,2013PhRvL.111m1101B,2013PhRvD..88d4026H,2014PhRvD..89j4029N,2014PhRvD..89j4021B,2014PhRvL.113i1104T,2014arXiv1412.3240T,2014arXiv1411.7975K}. Specifically, it has been shown that detections of its {\it dominant} post-merger GW frequency $f_{\mathrm{peak}}$ would strongly constrain the radius and the maximum mass of nonrotating NSs~\cite{2012PhRvL.108a1101B,2012PhRvD..86f3001B,2013PhRvD..88d4026H,2014PhRvD..90b3002B,2014PhRvD..90f2004C}. Additionally, there exist potentially detectable \textit{secondary peaks} at lower frequencies~\cite{1994PhRvD..50.6247Z,2002PhRvD..65j3005O,2002PThPh.107..265S,2005PhRvD..71h4021S,2006PhRvD..73f4027S,2007PhRvL..99l1102O,2008PhRvD..78h4033B,2009PhRvD..80f4037K,2011MNRAS.418..427S,2011PhRvD..83l4008H,2011PhRvL.107e1102S,2012PhRvL.108a1101B,2012PhRvD..86f3001B,2013PhRvD..88d4026H,2014PhRvD..89j4021B,2014PhRvL.113i1104T,2014arXiv1412.3240T,2014arXiv1411.7975K,2014PhRvD..90b3002B,2014PhRvD..90f2004C,2014PhRvX...4d1004M}, which could improve the constraints on 
NS properties if the mechanism by which they appear is 
understood.

In this study we find that \textit{two different mechanisms} are at work for producing the low-frequency secondary peaks in the GW spectrum at frequencies below the main peak of the post-merger phase. Specifically, we show that apart from a \textit{nonlinear combination frequency} \cite{2011MNRAS.418..427S}, there exists also a distinct mechanism that generates a secondary GW peak by the \textit{rotating pattern of a deformation of spiral shape}. This deformation is initially produced at the time of merging and is then sustained for a few rotational periods. The consideration of the two different mechanisms leads to a unified classification scheme for the post-merger dynamics and GW emission. For high-mass binaries (relative to the threshold mass to prompt black-hole collapse), the nonlinear combination frequency dominates, while for low-mass binaries it is the spiral deformation that produces the strongest secondary peak. Both are simultaneously present and can produce peaks of comparable strength for 
intermediate binary masses. 
Hence, the secondary peaks  cannot simply be explained by the dynamics of the double-core structure alone, which forms the inner part of the remnant, as in~\cite{2014PhRvL.113i1104T,2014arXiv1412.3240T,2014arXiv1411.7975K}. Identifying the importance of the rotating deformation, we explain for the first time the existence of a low-frequency modulation appearing in addition to the quasi-radial oscillation mode in several physical quantities characterizing the remnant. 

After clarifying the nature of the secondary peaks in the GW spectrum, we find mass-dependent empirical relations for the different peaks as a function of NS compactness. We rule out the existence of a universal, mass-independent and EoS independent relation for the secondary peaks, a relation recently proposed in~\cite{2014PhRvL.113i1104T,2014arXiv1412.3240T}. Such a relation can only be found within a small sample of EoSs for a very limited mass range (not the same for all EoSs). Our new relations can be used to optimize the constraints on the EoS in the case of the simultaneous detection of several post-merger frequencies. The understanding of the most prominent features of the GW spectrum is a prerequisite for constructing GW templates for the post-merger phase, which could enhance the detection prospects compared to unmodelled searches~\cite{2014PhRvD..90f2004C,2014PhRvX...4d1004M} for the Advanced LIGO and Advanced Virgo detectors and their discussed upgrades~\cite{Hild:2011np,2014RvMP...86..121A,
2015PhRvD..91f2005M}. For the planned Einstein Telescope~\cite{2010CQGra..27a5003H} direct detections of secondary peaks are a viable prospect~\cite{2014PhRvX...4d1004M,2014PhRvD..90f2004C,2014PhRvL.113i1104T,2014arXiv1412.3240T}.

\section{Nature of secondary GW peaks}
We investigate mergers of equal-mass, intrinsically non-spinning NSs with a 3D relativistic smoothed particle hydrodynamics (SPH) code, which imposes the conformal flatness condition on the spatial metric~\cite{1980grg..conf...23I,1996PhRvD..54.1317W} to solve Einstein's field equations and incorporates energy and angular momentum losses by a GW backreaction scheme~\cite{2003PhRvD..68h4001F,2007A&A...467..395O} (see~\cite{2002PhRvD..65j3005O,2007A&A...467..395O,2010PhRvD..82h4043B,2012PhRvL.108a1101B,2012PhRvD..86f3001B} for details on the code, the setup, resolution tests and model uncertainties). Comparisons to other numerical setups and also models with an approximate consideration of neutrino effects show an agreement in determining the post-merger spectrum within a few per cent in the peak frequencies~\cite{2011PhRvL.107e1102S,2012PhRvL.108a1101B,2012PhRvD..86f3001B,2013PhRvD..88d4026H,2014PhRvL.113i1104T,2014arXiv1412.3240T,2014arXiv1411.7975K}. Magnetic field effects are negligible for 
not too high initial field strengths~\cite{2011PhRvD..83d4014G}. We explore a representative sample of ten 
microphysical, fully temperature-dependent equations of state (EoSs) (see Table~I in~\cite{2014PhRvD..90b3002B} and Fig.~\ref{fig:class} in this work for the mass-radius relations of non-rotating NSs of these EoSs) and consider total binary masses $M_{\mathrm{tot}}$ between 2.4~$M_{\odot}$ and 3.0~$M_\odot$. In this work we consider only NSs with an initially irrotational velocity profile because known spin periods in observed NS binaries are slow compared to their orbital motion (see e.g.~\cite{2008LRR....11....8L}), and simulations with initial intrinsic NS spin suggest an impact on the post-merger features of the GW signal only for very fast spins~\cite{2007PhRvL..99l1102O,2014PhRvD..89j4021B,2014arXiv1411.7975K}.

\begin{figure}
\includegraphics[width=8.9cm]{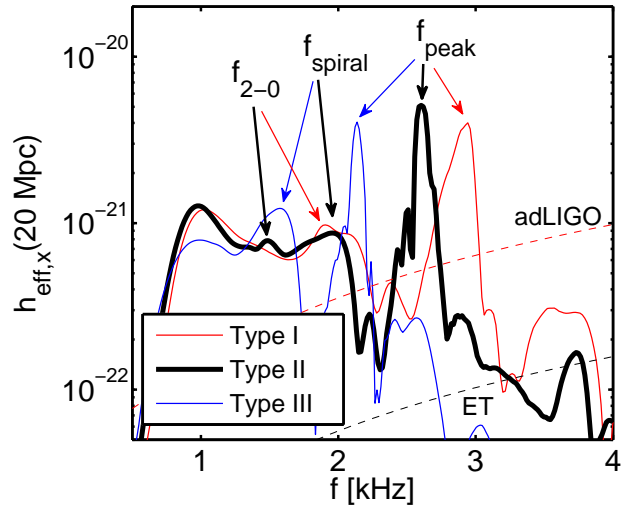}
\caption{\label{fig:spec}GW spectra of 1.35-1.35~$M_\odot$ mergers with the DD2~\cite{2010NuPhA.837..210H,2010PhRvC..81a5803T} (black), NL3~\cite{1997PhRvC..55..540L,2010NuPhA.837..210H} (blue) and LS220~\cite{1991NuPhA.535..331L}  (red) EoS (cross polarization along the polar axis at a reference distance of 20~Mpc). Dashed lines show the anticipated unity SNR sensitivity curves of Advanced LIGO~\cite{2010CQGra..27h4006H} (red) and of the Einstein Telescope~\cite{2010CQGra..27a5003H} (black).}
\end{figure}

First, we focus on a reference model for the moderately stiff DD2 EoS~\cite{2010NuPhA.837..210H,2010PhRvC..81a5803T} with an intermediate binary mass of $M_{\mathrm{tot}}=2.7~M_\odot$. Figure~\ref{fig:spec} shows the x-polarization of the  effective amplitude $h_{\mathrm{eff,x}}=\tilde{h}_\mathrm{x}(f)\cdot f$ (with $\tilde{h}_\mathrm{x}$ being the Fourier transform of the waveform $h_{\mathrm{x}}$) vs. frequency $f$ (reference model in black). Besides the dominant $f_{\rm peak}$ frequency~\footnote{Other references in the literature call the dominant frequency $f_2$.}, there are two secondary peaks at lower frequencies ($f_{2-0}$ and $f_\mathrm{spiral}$) with  comparable signal-to-noise ratio. Both are generated in the post-merger phase, which can be seen by choosing a time window covering only the post-merger phase for computing the GW spectrum.

The secondary peak shown as $f_{2-0}$ is a nonlinear combination frequency between the dominant quadrupolar $f_{\rm peak}$ oscillation and the quasi-radial oscillation of the remnant, as described in~\cite{2011MNRAS.418..427S}. We confirm this by performing additional simulations, after adding a quasi-radial density perturbation to the remnant at late times. The frequency $f_0$ of the strongly excited quasi-radial oscillation is determined by a Fourier analysis of the time-evolution of the density or central lapse function and coincides with the frequency difference $f_\mathrm{peak}-f_{2-0}$. As in~\cite{2011MNRAS.418..427S}, the extracted eigenfunction at $f_0$ confirms the quasi-radial nature.

\begin{figure*}
\includegraphics[width=8.9cm]{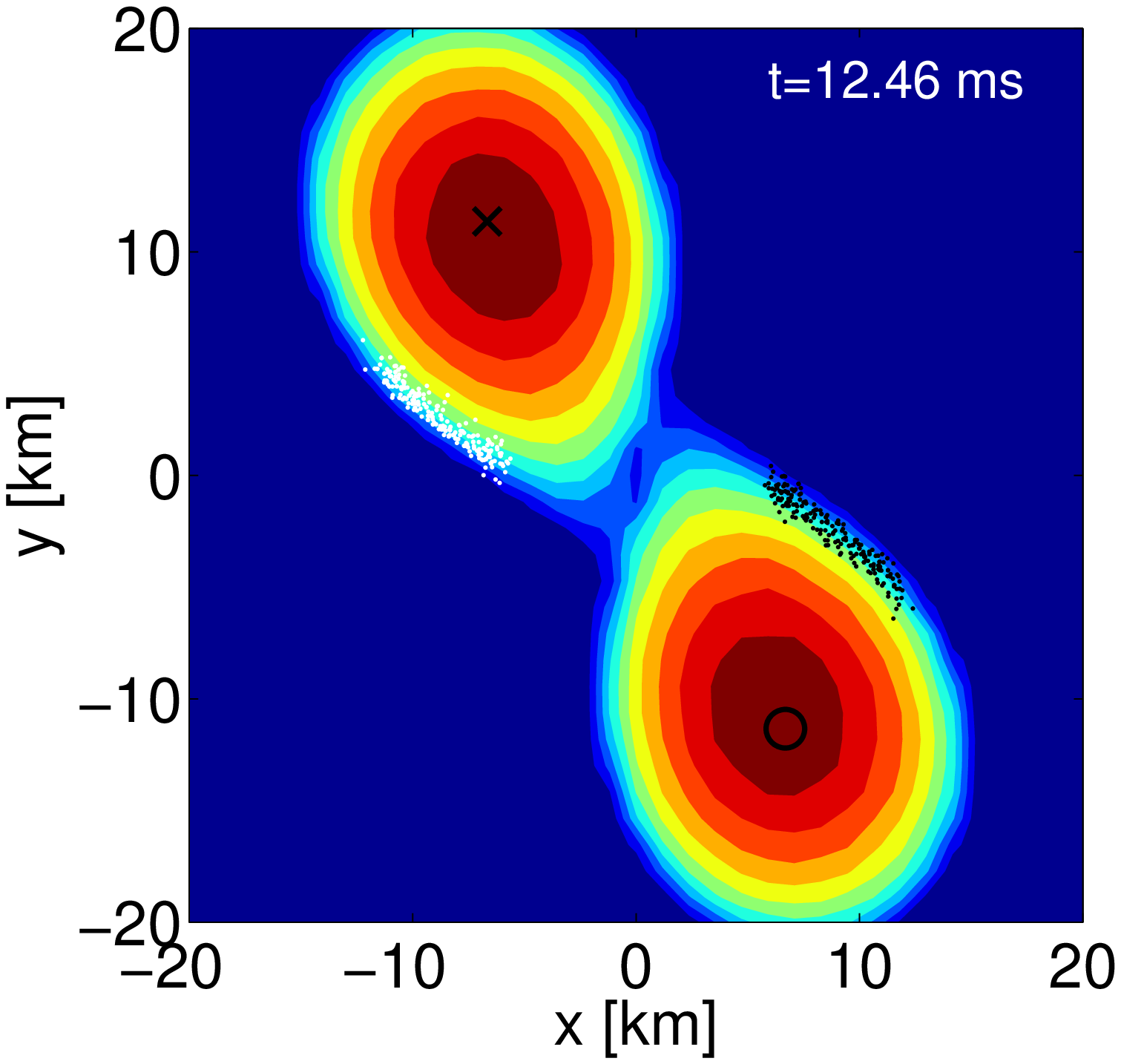} \includegraphics[width=8.9cm]{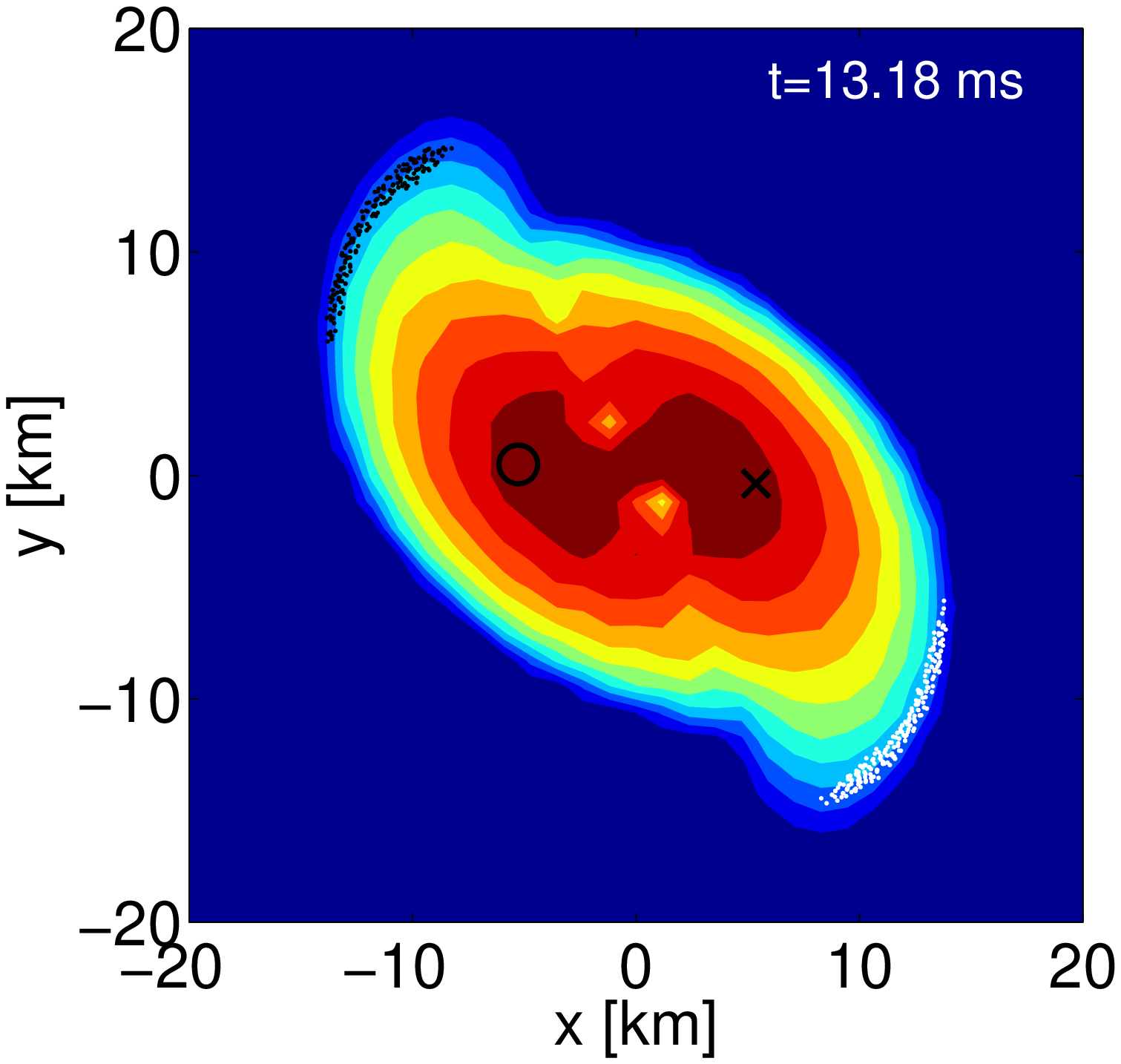}
\includegraphics[width=8.9cm]{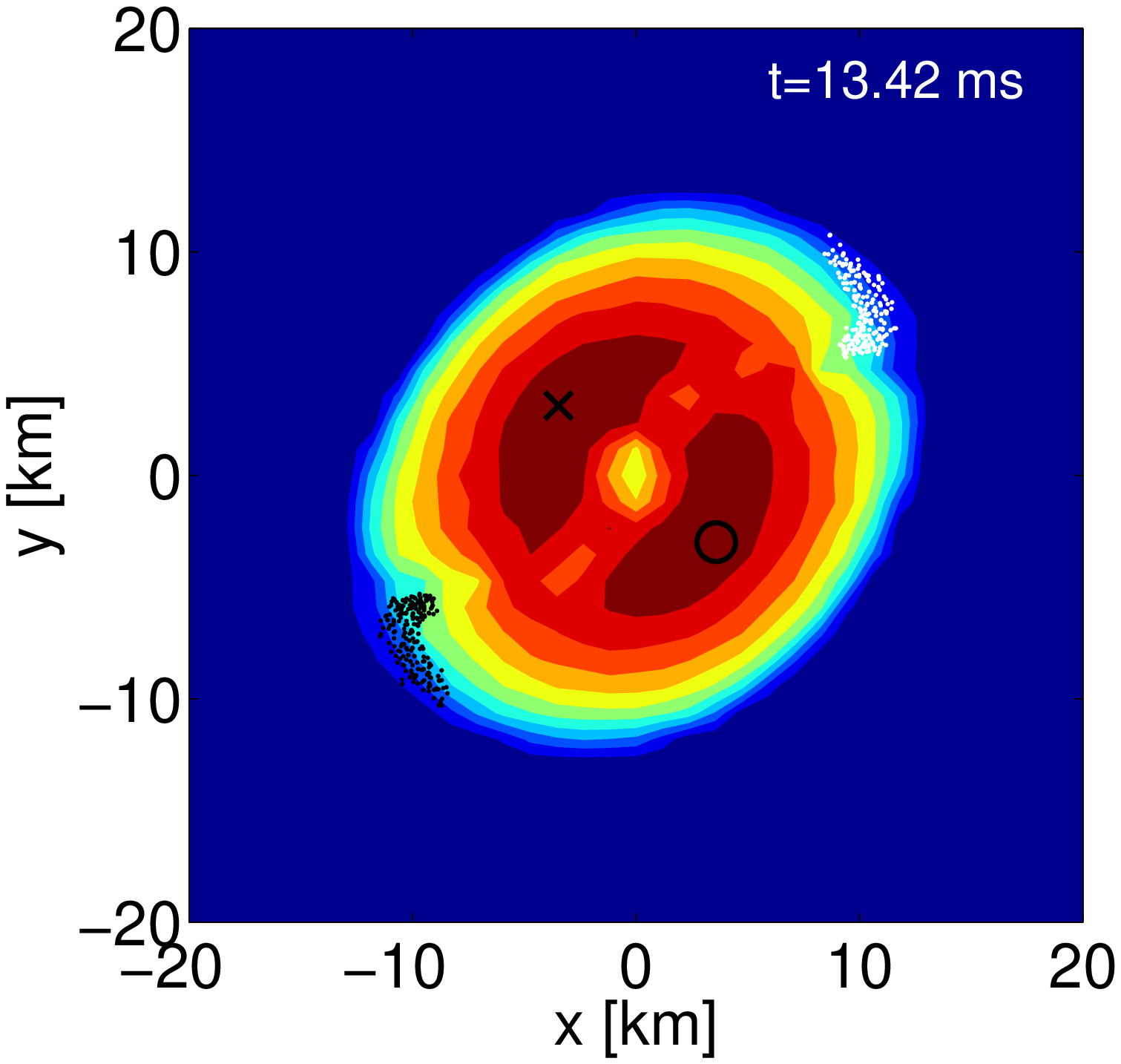} \includegraphics[width=8.9cm]{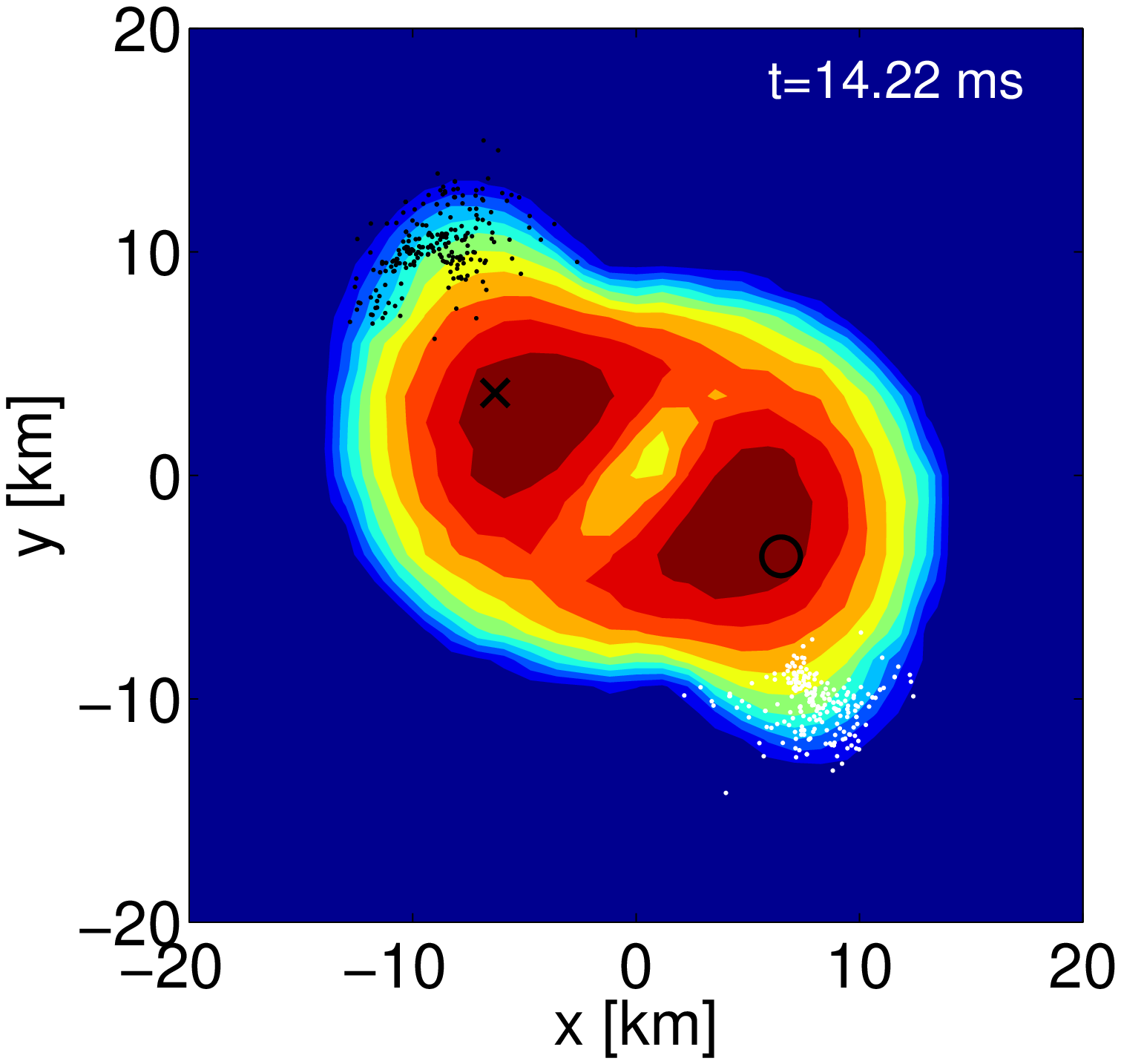}
\caption{\label{fig:snap}Rest-mass density evolution in the equatorial plane for the 1.35-1.35~$M_\odot$ merger with the DD2 EoS (rotation counter-clockwise). (The rest-mass density is shown with a variable linear scale relative to $\rho_\mathrm{max}$. A low number of contour levels is chosen for illustrative reasons; the underlying simulation data is smoother than it appears with the chosen color coding.) Black and white dots trace the positions of selected fluid elements of the antipodal bulges, which within approximately one millisecond complete one orbit (compare times of the right panels). The orbital motion of this pattern of spiral deformation produces the $f_{\rm spiral}$ peak in the GW spectrum at $2/(1~\mathrm{ms})$ (Fig. 1). The cross and the circle mark the double cores, which rotate significanty faster than the antipodal bulges represented by the dots (compare times of the different panels).}
\end{figure*}

The secondary $f_\mathrm{spiral}$ peak is produced by a strong deformation initiated at the time of merging, the pattern of which then rotates (in the inertial frame) slower than the inner remnant and lasts for a few rotational periods, while diminishing in amplitude. Figure~\ref{fig:snap} shows the density evolution in the equatorial plane, in which one can clearly identify the two antipodal bulges of the spiral pattern, which rotate slower than the central parts of the remnant. In this early phase the inner remnant is still composed of two dense cores rotating around each other (this is the nonlinear generalization of an $m=2$  quadrupole oscillation producing the dominant  $f_{peak}$). Extracting the rotational motion of the antipodal bulges in our simulations, we indeed find that their frequency equals $f_{\mathrm{spiral}}/2$ producing gravitational waves at $f_{\mathrm{spiral}}$ (compare the times in the right panels in Fig.~\ref{fig:snap}; recall the factor two in the frequency of the GW signal 
compared to the orbital frequency of orbiting point particles). In Fig.~\ref{fig:snap} the antipodal bulges are illustrated by selected fluid elements (tracers), which are shown as black and white dots, while the positions of the individual centers of the double cores are marked by a cross and a circle. (We define the centers of mass of the double cores by computing the centers of mass of the innermost 1000 SPH particles of the respective initial NSs and then following their time evolution.) While in the right panels the antipodal bulges completed approximately one orbit within one millisecond ($\approx \frac{2}{f_{\mathrm{spiral}}}$), the double cores moved further ahead, i.e. with a significantly higher orbital frequency. Examining the GW spectrum and considering different time intervals, we find that the presence of the $f_\mathrm{spiral}$ peak agrees with the appearance and duration of the spiral deformation of the remnant.

In the upper right panel of Fig. 2, the spiral deformation can be seen to initially reach deep inside the remnant. We approximately determine the amount of matter which belongs to the two antipodal bulges that are rotating slower compared to the double cores. This matter amounts to several tenths of $M_\odot$ and is thus sufficient to explain the strength of the $f_\mathrm{spiral}$ GW peak. In addition, we find that the $f_\mathrm{spiral}$ GW peak can be roughly reproduced in a toy model, where the two bulges orbit as point particles around the central double-core structure for a duration of a few milliseconds. Note that this toy model differs significantly from the one in~\cite{2014arXiv1412.3240T}, which considers only the two cores to be contributing to the GW signal and considers only a single instantaneous orbital frequency of the system.

\begin{figure}
\includegraphics[width=8.9cm]{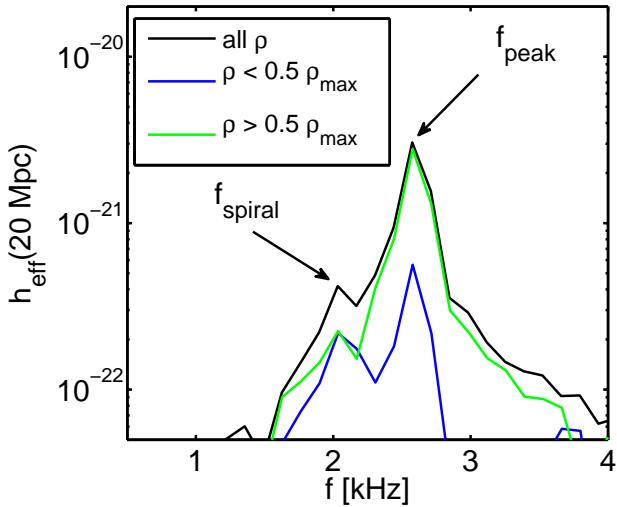}
\caption{\label{fig:rhospec}Early post-merger spectrum of 1.35-1.35~$M_\odot$ merger with the DD2 EoS (black) and GW spectra computed for fluid elements with densities above or below 50\% of the maximum density.}
\end{figure}

Furthermore, we take advantage of the quadrupole formalism to compute GW spectra considering only certain parts of the remnant, which are defined by using either a density or a spatial cut-off. In Fig.~\ref{fig:rhospec} we demonstrate that the dominant quadrupole  $f_{peak}$ frequency is generated mainly by those regions of the remnant which encompass densities \textit{exceeding} 50\% of the instantaneous maximum density $\rho_\mathrm{max}$. In contrast, most power of the $f_\mathrm{spiral}$ peak originates from densities \textit{below} $0.5 \rho_\mathrm{max}$, which corresponds to the outer parts of the remnant, where the two bulges form (see Fig.~\ref{fig:snap}). Similar conclusions are reached when a spatial cut-off instead of a density cut-off is used.

\begin{figure}
\includegraphics[width=8.8cm]{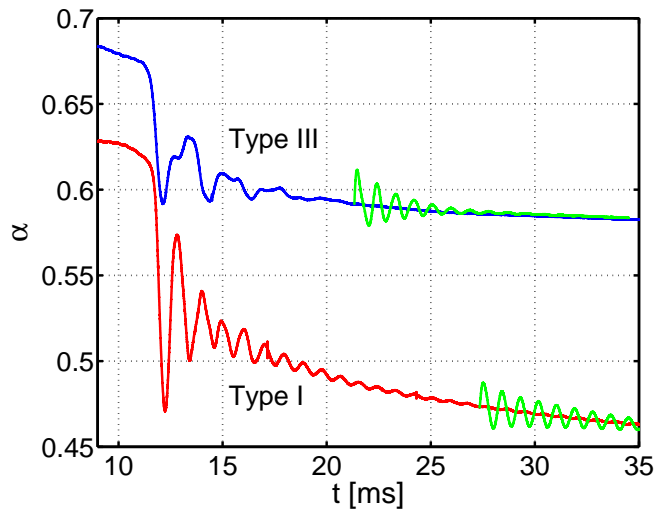}
\caption{\label{fig:lapse}Evolution of the central lapse function for 1.35-1.35~$M_\odot$ mergers with the NL3 (blue) and LS220 (red) EoS (time-shifted to the same merging time). The evolution of models with an added quasi-radial perturbation is shown (green). For Type II and Type III mergers, like the model with the NL3 EoS shown here, one recognizes a low-frequency oscillation in addition to the quasi-radial mode, which is also strongly excited in the perturbed model at late times.}
\end{figure}

In models where $f_\mathrm{spiral}$ dominates over $f_{2-0}$ (see
our classification below), the presence of the two rotating antipodal bulges explains the appearance of a particular \textit{low-frequency modulation} with $f_\mathrm{peak}-f_\mathrm{spiral}$ seen in the time evolution of the central lapse function (blue curve in Fig.~\ref{fig:lapse}), of  $\rho_\mathrm{max}$, of the size of the central remnant and of the separation between the two cores. The same modulation occurs as a beat frequency in the time evolution of the GW amplitude (see e.g. Fig.~1 in~\cite{2007PhRvL..99l1102O}). The low-frequency modulation with $f_\mathrm{peak}-f_\mathrm{spiral}$ can coexist with the quasi-radial oscillation, which has a higher frequency $f_{0}$. 

The above modulation frequency is associated with the rotating bulges and is explained as follows: the  central remnant forms an elongated structure, around which the two bulges rotate, lagging behind. The characteristics of the remnant are modulated depending on the orientation of the antipodal bulges with respect to the double cores: the compactness is smaller, the central lapse function larger and the GW amplitude maximal when the bulges and the cores are aligned (lower right panel in Fig.~\ref{fig:snap}). Instead, when the bulges and the cores are orthogonal to each other (lower left panel), the compactness is largest, the central lapse function smaller and the GW amplitude small. 

In a frame corotating with the central remnant (i.e. a frame rotating with frequency $\sim f_\mathrm{peak}/2$ w.r.t. the inertial frame) the two bulges counter-rotate with a frequency of $\Delta f=(f_{\mathrm{peak}}-f_\mathrm{spiral})/2$ which corresponds to the $f_{\mathrm{peak}}-f_\mathrm{spiral}$ modulation in various quantities (notice the $\pi$-symmetry of the system, which compensates the factor $1/2$).

Our simulations also show that the rotational frequency of the centers of
the double-core structure, although not constant,  significantly exceeds
$f_{\mathrm{spiral}}/2$ at any time. Hence, variations in the angular frequency of the double-core structure alone are not sufficient to interpret the secondary peaks, and the simultaneous presence of both  $f_\mathrm{peak}$ and $f_{\mathrm{spiral}}$ cannot simply be attributed to a single instantaneous angular frequency of the system, as suggested in~\cite{2014PhRvL.113i1104T,2014arXiv1412.3240T,2014arXiv1411.7975K}. We note that the simultaneous presence of two (orbital) frequencies  (of the double-core structure and of the antipodal bulges) naturally explains strong time variations of the instantaneous GW frequency as seen in many simulations, e.g.~\cite{2013PhRvD..88d4026H,2014arXiv1411.7975K}.

Also, a single, initially strongly varying instantaneous frequency as explanation for the peaks in the spectrum is incompatible with the fact that the most pronounced peak in the spectrum occurs already at early times. Initially, the instantaneous frequency strongly oscillates around the frequency of the dominant peak. According to the picture of~\cite{2014arXiv1411.7975K}, peaks form at frequencies at which the instantaneous frequency spends most time, i.e. at the extrema of the instantaneous frequency. In this case, one would thus not expect that the dominant $f_\mathrm{peak}$ is strong at early times. In contrast, in our simulations the dominant $f_\mathrm{peak}$ has a substantial strength if one considers only the first few milliseconds.

\section{Classification of post-merger dynamics and GW emission} 
We have applied the above analysis tools (GW spectra, determination of $f_0$ from perturbed models, rotational frequency of the antipodal bulges and of the double cores, GW spectra with different cut-off densities) for a number of representative models, varying the binary mass and stiffness of the EoS. The results fully confirm the generic picture described above. Considering this larger set of models, we identify, based on the relative strength between $f_{2-0}$ and $f_\mathrm{spiral}$, three different types of post-merger dynamics and GW spectra for remnants which survive for more than several milliseconds.

\begin{itemize}
\item \textit{Type I:} When the total binary mass $M_\mathrm{tot}$ is not too far from the threshold mass for prompt quasi-radial collapse of the remnant for a given EoS~\cite{2013PhRvL.111m1101B}, the evolution of the central lapse function (and of $\rho_\mathrm{max}$) is dominated by a very strong \textit{quasi-radial oscillation} of the remnant, see lower curve in Fig.~\ref{fig:lapse}. For such models the two initial NSs are more centrally condensed and they merge with higher impact velocity (Fig.~3 in~\cite{2013ApJ...773...78B}). Because of the strongly excited quasi-radial oscillation, $f_{2-0}$ is the strongest secondary peak in the GW spectrum, while $f_{\rm spiral}$ is much weaker, likely because for more compact NSs the formation of the spiral pattern is less pronounced. There can be a partial overlap between $f_{2-0}$ and $f_{\rm spiral}$, see red curve in Fig.~\ref{fig:spec}.

\item \textit{Type II:}  For intermediate total binary masses, $f_{2-0}$ and $f_\mathrm{spiral}$ have a \textit{comparable strength} in the GW spectrum and the two types of secondary peaks are \textit{well separated} (see black curve Fig. 1). This is the generic type, which we discussed earlier in our reference model. In characteristic quantities such as the central lapse function the quasi-radial oscillation frequency $f_0$ as well as the low-frequency modulation with $f_{\mathrm{peak}}-f_\mathrm{spiral}$ are clearly noticable.

\item \textit{Type III:} When the total binary mass $M_\mathrm{tot}$ is significantly below the threshold
mass for quasi-radial collapse, the time evolution of the central lapse function
(as well as of $\rho_\mathrm{max}$ and of the radius of the remnant) is dominated by the  $f_{\mathrm{peak}}-f_\mathrm{spiral}$ modulation that we explained in the previous section as a result of the rotating spiral pattern with the two antipodal bulges. In the evolution of the central lapse function this modulation typically has a smaller amplitude than Type I variations (Fig.~\ref{fig:lapse}).
The quasi-radial oscillation $f_0$ is also present, but with much smaller amplitude than the dominant modulation because the smaller NS compactness implies a smaller impact velocity and thus a weaker excitation of the quasi-radial oscillation. The smaller NS compactness also allows for a stronger spiral deformation.  Consequently, the dominant secondary peak in the GW spectrum is $f_\mathrm{spiral}$, while  $f_{2-0}$ is either very weak or hidden inside the background (see blue curve in Fig.~\ref{fig:spec}). 

\end{itemize}

\begin{figure}
\includegraphics[width=8.9cm]{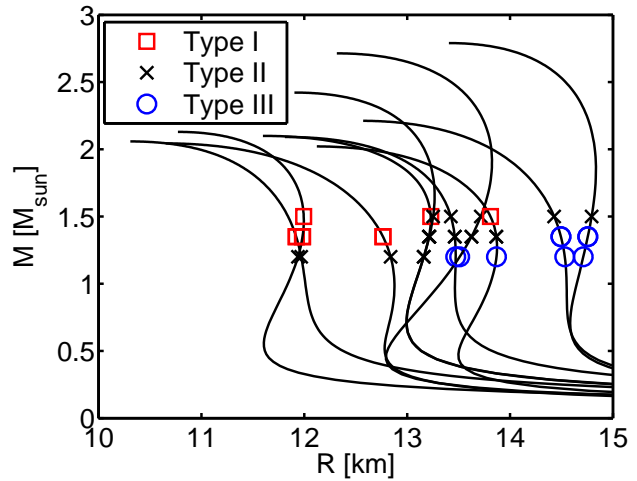}
\caption{\label{fig:class}Different types of post-merger dynamics and GW emission of the different merger models visualized by the mass-radius relations of non-rotating NSs of the EoSs considered in this work. The outcome of a given calculation with $M_{\mathrm{tot}}$ is shown as symbol at $M_{\mathrm{tot}}/2$ plotted on the mass-radius relation of the EoS employed in the simulation. Red squares indicate Type I, black crosses stand for Type II, and blue cirlces mark Type III. See text for definitions of different types of post-merger dynamics and GW emission.}
\end{figure}

For a given EoS there is a continuous transition between the different types of post-merger dynamics depending on the total binary mass. Types I and III are the limiting cases of the more generic Type II. Notice that since the threshold for quasi-radial collapse is EoS-dependent~\cite{2013PhRvL.111m1101B}, the different types cover a different mass range for each EoS.  For a total binary mass of $M_\mathrm{tot}=2.7~M_\odot$ all three types are possible depending on the EoS, where very soft EoSs yield Type I mergers and very stiff EoSs lead to Type III dynamics (see GW spetra in Fig.~\ref{fig:spec}). This is also shown in Fig.~\ref{fig:class}, which provides an overview of the types of the different models considered in this study. The type of a given simulation is indicated by a symbol plotted at the mass of the individual inspiralling stars on the mass-radius relation of the EoS which was used in the calculation (e.g. the results of the 1.2-1.2~$M_\odot$ binaries are displayed at the radii of NSs with 1.
2~$M_\odot$). As described, there is a continuous transition between the different classes, which is why one should consider the transitions between the different types in Fig.~\ref{fig:class} as being tentative and a slightly different classification may be possible at the borders between the different types. More quantitative definitions of the types may be useful in the future. Still, one can clearly identify a diagonal band of Type II mergers for intermediate binary masses, and also the binary setups leading to the limiting cases of Type I or Type III are seen to form roughly diagonal bands.

For $2.4~M_\odot \le M_\mathrm{tot}\le 3.0~M_\odot$ we find that $f_\mathrm{spiral}$
typically ranges between $f_\mathrm{peak}-0.5$~kHz and $f_{\mathrm{peak}}-0.9$~kHz,
while $f_{2-0}$ ranges between $f_{\mathrm{peak}}-0.9$~kHz and $f_{\mathrm{peak}}-1.3$~kHz.
This property will be useful for identifying either $f_{2-0}$ or $f_\mathrm{spiral}$
(or both) in future GW observations. Furthermore, we find that $f_\mathrm{peak}-f_{2-0}(=f_0)$ \textit{decreases} with increasing $M_\mathrm{tot}$ in all models for which $f_{2-0}$ is clearly present, in agreement with the fact that the quasi-radial frequency decreases near the threshold to collapse. This observation may be useful to estimate the proximity to prompt gravitational collapse. Very near the threshold one thus may expect $f_{2-0}\rightarrow f_\mathrm{peak}$. In contrast, $f_\mathrm{peak}-f_{\rm spiral}$  typically \textit{increases} with increasing $M_\mathrm{tot}$, and above the threshold to collapse a spiral pattern during the dynamical collapse could still produce a weak peak in the GW spectrum, as in~\cite{2010PhRvL.104n1101K}.

\begin{figure}
\includegraphics[width=8.9cm]{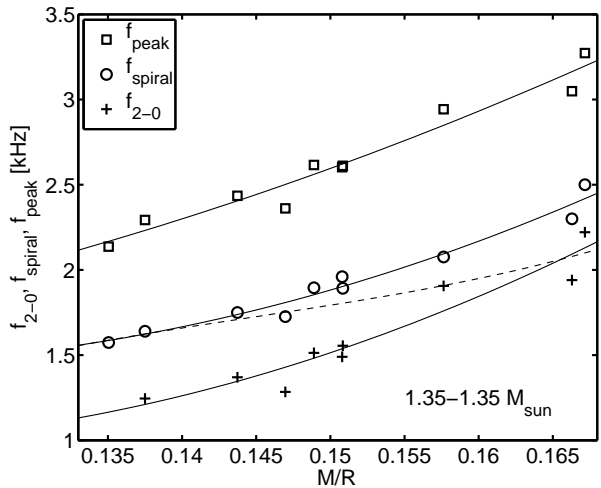}
\caption{\label{fig:freqa}$f_\mathrm{peak}$, $f_\mathrm{spiral}$ and $f_{2-0}$ for mergers with ten different EoSs and $M_\mathrm{tot}$=2.7~$M_\odot$ vs. the compactness $M/R$ for nonrotating, single NSs. Solid lines show empirical relations. The dashed line is taken from~\cite{2014arXiv1412.3240T} (see text for explanations).}
\end{figure}
\begin{figure}
\includegraphics[width=8.9cm]{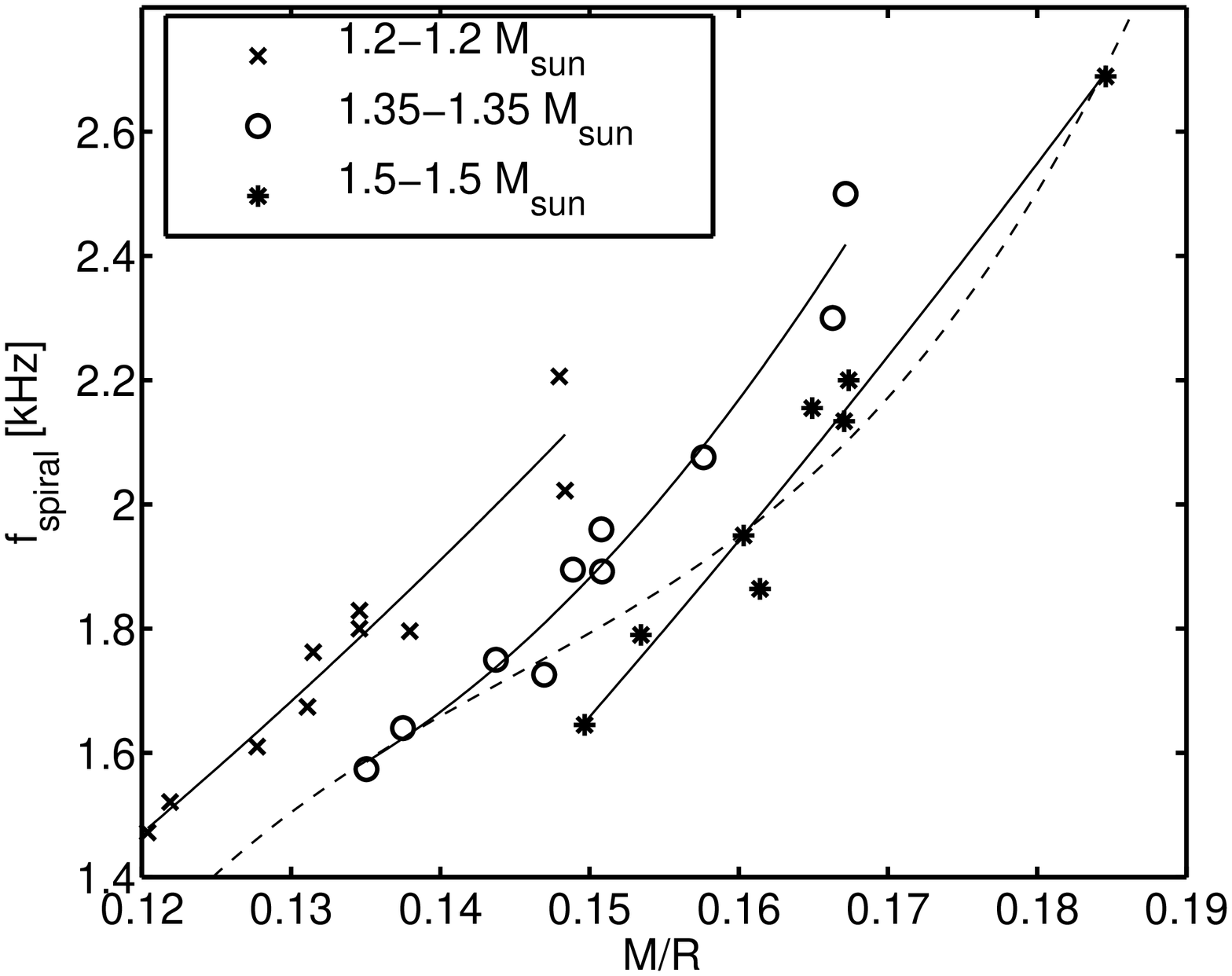}
\caption{\label{fig:freqb}$f_\mathrm{spiral}$ vs. the compactness, but for different binary masses. Solid lines show empirical relations. The dashed line is taken from~\cite{2014arXiv1412.3240T} (see text for explanations).}
\end{figure}
\begin{figure}
\includegraphics[width=8.9cm]{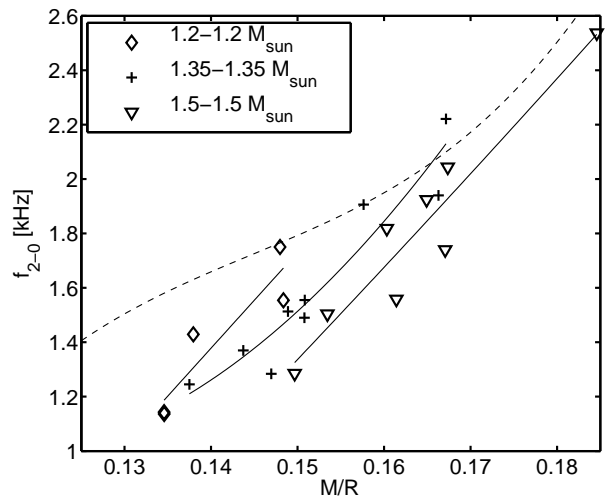}
\caption{\label{fig:freqc}$f_{2-0}$ vs. the compactness, but for different binary masses. Solid lines show empirical relations. The dashed line is taken from~\cite{2014arXiv1412.3240T} (see text for explanations).}
\end{figure}

\section{Empirical relations for dominant and secondary peak frequencies}
For our sample of EoSs Fig.~\ref{fig:freqa} shows $f_\mathrm{peak}$, $f_\mathrm{spiral}$ and $f_{2-0}$ as a function of the compactness $ M/R$ of the \textit{nonspinning, individual NSs} (at infinite separation) for $M_{\mathrm{tot}}=2.7~M_\odot$ (with the compactness in units of $c=G=1$). We find strong correlations that can be described by the following quadratic fits: 
\begin{eqnarray}
f_\mathrm{peak} [\rm kHz]&=&199 (M/R)^2 -28.1(M/R) +2.33 , \label{eq:fpeak} \\  
f_\mathrm{spiral} [\rm kHz] &=& 358 (M/R)^2-82.1 (M/R) + 6.16, \label{eq:fspiral} \\ 
f_{2-0} [\rm kHz] &=& 392 (M/R)^2 - 88.3 (M/R) +5.95. \label{eq:f20}
\end{eqnarray}
The maximum deviations of the data used for these fits are 140~Hz, 86~Hz and 153~Hz for $f_\mathrm{peak}$, $f_\mathrm{spiral}$ and $f_{2-0}$, respectively. If the compactness is determined from a measured frequency by inverting Eqs.~\eqref{eq:fpeak}-\eqref{eq:f20}, these maximum deviations imply errors of 3\%, 3\% and 4\% in the compactess for $f_\mathrm{peak}$, $f_\mathrm{spiral}$ and $f_{2-0}$, respectively. (Note that the slope in the relation for $f_\mathrm{spiral}$ is somewhat flatter.) Thus, the accuracy of these empirical relations is similarly good for $f_\mathrm{peak}$ and $f_\mathrm{spiral}$ and only slightly worse for $f_{2-0}$. Even if one assumes an error of 10\% uncertainty in the frequency determinations from an actual GW observation, (which at least for $f_\mathrm{peak}$ is too pessimistic~\cite{2014PhRvD..90f2004C}) this would add a 6\%, 8\% and 5\% per cent error in the compactness determination. Similar empirical relations hold, with varying accuracy, for each total binary mass. As already 
pointed out and explained for $f_\mathrm{peak}$ 
in~\cite{2012PhRvD..86f3001B}, even better empirical relations are obtained for 1.35-1.35~$M_\odot$ binaries when the above three frequencies are examined as function of the radius of a nonrotating NS with 1.6~$M_{\odot}$. Notice that all three empirical relations follow similar trends.

Because the frequency of the $f_\mathrm{spiral}$ peak agrees with the rotational frequency of the antipodal bulges near the surface of the remnant, a scaling of $f_\mathrm{spiral}$  with the compactness is not unexpected, which explains the similarities with the behavior of $f_\mathrm{peak}$. In fact, for the models with $M_{\mathrm{tot}}=2.7~M_\odot$ one finds very tight relations between $f_\mathrm{spiral}$ and $f_\mathrm{peak}$, and between $f_{2-0}$ and  $f_\mathrm{peak}$.

A measurement of $f_\mathrm{peak}$   is already sufficient to accurately constrain the radius of nonrotating NSs (and thus the EoS)~\cite{2012PhRvL.108a1101B,2012PhRvD..86f3001B,2014PhRvD..90b3002B} if the total mass has been obtained accurately from the inspiral signal, as it is likely to be the case for distances within which  $f_\mathrm{peak}$ has the required signal-to-noise ratio to be detectable with second-generation interferometers (e.g.~\cite{2014ApJ...784..119R,2014PhRvD..90f2004C}).
A detection of the weaker secondary peaks, with similar $M/R$ dependence as $f_\mathrm{peak}$, could further optimize the constraints on the EoS.

Figure~\ref{fig:freqb} displays $f_\mathrm{spiral}$ as a function of the compactness $M/R$ of the nonspinning, individual NSs for different EoSs and for different $M_\mathrm{tot}$ in the range 2.4-3.0~$M_{\odot}$ (the most likely range of total binary masses~\cite{2012ARNPS..62..485L}). The same figure,  but for $f_{2-0}$, is shown in Fig.~\ref{fig:freqc}. Especially for $f_\mathrm{spiral}$ we discover that there exist tight relations between the compactness and the secondary frequencies for fixed binary masses.  

In \cite{2014PhRvL.113i1104T,2014arXiv1412.3240T}~the existence of a single, universal, mass-independent relation (dashed line in Figs.~\ref{fig:freqa} to~\ref{fig:freqc}) between the frequency of the strongest secondary GW peak (denoted there as $f_1$) and $M/R$ was proposed (there was no distinction of two different secondary peaks, as we find here). However, this result was based on using a limited set of five EoSs of soft or moderate stiffness (with corresponding maximum masses of nonrotating NS only up to 2.2~$M_{\odot}$)  as well as on different chosen mass ranges for each EoS with a spread of only 0.2~$M_{\odot}$ in the total binary mass. 

In contrast to~\cite{2014PhRvL.113i1104T,2014arXiv1412.3240T}, within our larger sample of ten EoSs (that includes stiff EoSs with maximum masses reaching up to  2.8~$M_{\odot}$) and for a more representative total binary mass range of 2.4-3.0~$M_{\odot}$ (same for all EoSs),   \textit{such a mass-independent, universal relation does not exist} (dashed curve in Figs.~\ref{fig:freqa} to~\ref{fig:freqc}). Figures~\ref{fig:freqb} and ~\ref{fig:freqc} show that there is a large spread of up to 500~Hz both in the $f_\mathrm{spiral}$
vs. $M/R$ relation and in the $f_{2-0}$ vs. $M/R$ relation for an expected range of total binary masses, and that the data cannot be described by a single function (dashed line). Even if one consistently chooses the strongest secondary peak among $f_\mathrm{spiral}$ and $f_{2-0}$ in each case, there does not exist a mass-independent universal relation with  compactness. The absence of a mass-independent, universal relation implies that the procedure for EoS constraints and binary mass determinations as described in~\cite{2014PhRvL.113i1104T,2014arXiv1412.3240T} is not valid. Notice also that in Fig.~\ref{fig:freqa}, the relation proposed in~\cite{2014PhRvL.113i1104T,2014arXiv1412.3240T} describes  either $f_\mathrm{spiral}$
at low compactness or $f_{2-0}$ at high compactness, which is consistent with the expectations from our classification scheme describing which of the secondary peaks dominates the GW spectrum. However, for intermediate values of the compactness the merger will be of Type II and both types of secondary peaks can be present with comparable amplitude, which further complicates the definition of a single $f_1$ frequency, as was assumed in~\cite{2014PhRvL.113i1104T,2014arXiv1412.3240T}. We also point out that detections of the secondary peaks may possibly be less accurate since the peaks are broader in comparison to the main peak and sometimes do not stand out clearly.

Instead of a universal mass-independent relation, we find that there exist useful empirical relations only for fixed binary masses (shown as thin line segments in Figs.~\ref{fig:freqb} and~\ref{fig:freqc}), such as the case shown in Fig.~\ref{fig:freqa}. For merger events sufficiently close to allow a detection of post-merger GW peaks an accurate determination of the binary masses is expected (e.g.~\cite{2014ApJ...784..119R}). With the upcoming GW detectors the dominant post-merger oscillation frequency has been shown to be detectable with high precision for mergers within several Mpc using a morphology-independent burst analysis~\cite{2014PhRvD..90f2004C}. Significant improvements in the detectability are possible for more sophisticated search algorithms, e.g. matched filtering, which, however, require detailed modelling of the expected signal features to which our present study should contribute by clarifying the relation between dominant and secondary GW peaks. Depending on the highly uncertain binary NS 
merger rate, a detection may succeed with Advanced LIGO/Virgo~\cite{2014PhRvD..90f2004C} or with discussed upgrades~\cite{Hild:2011np,2014RvMP...86..121A,2015PhRvD..91f2005M} (the latter may yield a sensitivity increase at high frequencies of a factor three). Since the strength of the secondary features is somewhat lower than that of the dominant peak (in relation to the anticipated instrument noise curves), direct detections of the secondary GW peaks can be expected for the planned Einstein Telescope~\cite{2010CQGra..27a5003H,2014PhRvX...4d1004M,2014PhRvD..90f2004C,2014PhRvL.113i1104T,2014arXiv1412.3240T}, unless the merger rate is on the more optimistic side as defined in~\cite{2010CQGra..27q3001A}, which may enable an earlier observation. The exact detection rate of secondary peaks is hard to quantify not only because of the uncertain merger rate, but also because of the varying strength and prominence of the secondary peaks depending on the exact model.

In this study we focus on equal-mass binaries. To assess the impact of the binary mass ratio, we perform additional simualtions for 1.3-1.4~$M_\odot$ binaries with the DD2 EoS~\cite{2010NuPhA.837..210H,2010PhRvC..81a5803T}, LS220 EoS~\cite{1991NuPhA.535..331L} and NL3 EoS~\cite{1997PhRvC..55..540L,2010NuPhA.837..210H}. The GW spectra in comparison to the symmetric case exhibit qualitatively the same features. The frequencies of the three different GW peaks are very similar to the equal-mass case with only small deviations of at most 2\%, 4\% and 6\% for $f_\mathrm{peak}$, $f_\mathrm{spiral}$ and $f_{2-0}$, respectively, compared to the results of the 1.35-1.35~$M_{\odot}$ binaries. Also the evolution of the central lapse function shows qualitatively the same behavior. This suggests that at least for moderately asymmetric binaries our classification scheme applies as well. It also implies that the frequency dependencies and their implications discussed above hold quantitatively with a good accuracy, although 
the secondary peaks show somewhat larger deviations, which may impede their use for EoS constraints if the mass ratio is not measured accurately from the GW inspiral signal.

\section{Summary and Outlook}
We present a unified picture of the post-merger dynamics and GW emission in binary NS mergers. Aside from the secondary GW peak produced by a nonlinear coupling between quadrupole and quasi-radial oscillations, we identify a new mechanism by which another secondary peak in the GW spectrum appears: it is caused by a spiral deformation, the pattern of which rotates slower with respect to the double-core structure of the inner remnant. Based on the presence and strength of these two secondary peaks, we introduce a classification scheme of the post-merger dynamics and of the GW spectrum. Moreover, the new mechanism allows for an explanation of low-frequency modulations seen in a number of physical characteristics of the remnant, such as the central lapse function, the maximum density, the separation between the two cores and the GW amplitude.

We find that for fixed total binary masses the frequencies of the secondary GW peaks show a tight relation with the compactness of nonrotating, individual NSs. These relations follow similar trends as the relation between the dominant peak frequency and the NS compactness (or, equivalently, the NS radius, as shown in~\cite{2012PhRvL.108a1101B,2012PhRvD..86f3001B}). For the dominant peak and for the secondary peak associated with the orbital motion of the antipodal bulges we find the relations between frequency and compactness to be similarly tight. We rule out the existence of a universal, mass-independent relation for secondary peaks as proposed in~\cite{2014PhRvL.113i1104T,2014arXiv1412.3240T}.

Identifying the type of a merger event and especially the determination of the quasi-radial frequency by $f_0=f_\mathrm{peak}-f_{2-0}$ can lead to an estimate of the threshold mass for black-hole collapse and thus to an estimate of the maximum mass of nonrotating NSs (see~\cite{2013PhRvL.111m1101B}). The insights from our classification scheme and from the frequency dependencies found here can provide a basis for constructing future GW template waveforms, increasing the changes for the observation of the post-merger GW emission~\cite{2014PhRvD..90f2004C}.

We will further investigate whether the frequency differences between the dominant and secondary peaks can clarify the nature of a detected secondary peak or if detailed comparisons between spectra of different binary masses are needed. We will also explore unequal-mass binaries (anticipating a strong impact of the mass-ratio on $f_\mathrm{spiral}$) and analyze the relevance of our classification scheme for the mass ejection and torus formation of NS mergers and for accompanying phenomena, such as r-process nucleosynthesis~\cite{1977ApJ...213..225L,1989Natur.340..126E}, electromagnetic counterparts~\cite{1998ApJ...507L..59L,2005astro.ph.10256K,2010MNRAS.406.2650M} and short gamma-ray bursts~\cite{1986ApJ...308L..43P,1989Natur.340..126E}.

\begin{acknowledgments}
We thank Matthias Hempel for providing EoS tables, and James Clark  and Thomas Janka for helpful discussions. A.B. is a Marie Curie Intra-European Fellow within the 7th European Community Framework Programme (IEF 331873). Partial support comes from ``NewCompStar'', COST Action MP1304. The computations were performed at the Rechenzentrum Garching of the Max-Planck-Gesellschaft, the Max Planck Institute for Astrophysics, and the Cyprus Institute under the LinkSCEEM/Cy-
Tera project.
\end{acknowledgments}


\end{document}